# Casimir and van der Waals forces: Advances and problems

G. L. Klimchitskaya, V. M. Mostepanenko

Central Astronomical Observatory at Pulkovo of the Russian Academy of Sciences, St.Petersburg, Russia,

Institute of Physics, Nanotechnology and Telecommunications, Peter the Great St. Petersburg Polytechnic University, St.Petersburg, Russia

Abstract: We review modern achievements and problems in physics of the van der Waals and Casimir forces which arise due to zero-point and thermal fluctuations of the electromagnetic field between closely spaced material surfaces. This subject attracted great experimental and theoretical attention during the last few years because the fluctuation-induced forces find a lot of applications in both fundamental physics and nanotechnology. After a short introduction to the subject, we describe main experimental and theoretical results obtained in the field during the last fifteen years. In the following presentation, we discuss some of the recent results by the authors and their collaborators which are of high promise for future developments. Specifically, we consider new features of the Casimir force acting between a gold sphere and an indium tin oxide plate, present the experimental and theoretical results on measuring the Casimir interaction between two gold surfaces by means of dynamic atomic force microscope, and outline first measurements of the Casimir interaction between magnetic surfaces and related theory. Special attention is devoted to the Casimir effect for graphene, which is the prospective material for microelectromechanical devices of next generations.

### 1. Introduction

It is common knowledge that the van der Waals [1] and Casimir [2] forces act between closely spaced, uncharged material surfaces made of different materials (metallic, dielectric or semiconductor). These forces are caused by the zero-point and thermal fluctuations of the electromagnetic field whose spectrum is altered by the presence of boundaries. The van der Waals and Casimir forces are of pure quantum origin. There are no such forces in classical electrodynamics at zero temperature. Taking into account that they arise not due to action of electric or magnetic fields, which mean values are zero, but due to the field dispersions, both forces are often called by a generic name of dispersion forces. In fact, there are no two different forces, van der Waals and Casimir. The van der Waals force is a subdivision of dispersion forces acting at very short separations up to a few nanometers, where the effect of relativistic retardation is very small and can be neglected. As to the Casimir force, it is a subdivision of dispersion forces which acts at larger separation distances, where the effect of relativistic retardation should be taken into account. It is evident that there is some transition region between the two kinds of dispersion forces.

The unified theory of the van der Waals and Casimir forces was developed by Lifshitz [3] in the middle of 1950s. According to the Lifshitz theory, the free energy and force of dispersion interaction between two parallel plates can be calculated if one knows the frequency-dependent dielectric permittivities and magnetic permeabilities of plate materials over wide frequency ranges. Using this theory many calculations were performed by different authors. Specifically, the previously obtained results by London [4] for the van der Waals force and by Casimir [5] were reproduced as the limiting cases of small and large separations between the plates.

During the second part of the twentieth century only a few experiments on measuring the Casimir force were performed (see review in [2]). The experimental results were compared with theoretical predictions of the Lifshitz theory, but only a qualitative agreement was achieved. The point is that there was no sufficiently precise laboratory technique for micromanipulation and preserving parallelity of the plates in vacuum at separations below a micrometer. Moreover, for dielectric materials used in several experiments performed at that period, the localized surface charges were an important background effect which plagued measurement of the Casimir force.

This situation has been changed in the end of 1990s when U. Mohideen [6] suggested to use an atomic force microscope (AFM) for measurements of the Casimir force. For this purpose the standard sharp tip used in all AFM until then was replaced with a metal-coated polystyrene sphere of about 100 μm radius. Only a bit later another prospective setup for measuring the Casimir force has been proposed using a micromachined oscillator [7]. Both new tools exploited the configuration of a large sphere situated in close proximity to a plane plate to measure the Casimir force. Historically the first, in 1997, there was a suggestion to measure the Casimir force between a plane plate and a spherical lens of about ten centimeter radius [8]. The use of this experimental scheme was associated with a lot of contradictory results. Later it was shown [9] that the measurement data obtained in this way are fundamentally flawed by the imperfections which are inevitably present on the surfaces of glass lenses of centimeter-size radii.

The present paper is organized as follows. In Sec.2 we briefly discuss modern achievements in the Casimir physics which became possible due to the use of new laboratory setups mentioned above. We also consider unexpected problems which originated from the comparison of the results of precise measurements with theoretical predictions of the Lifshitz theory. Section 3 is devoted to the new features of the Casimir force acting between a gold-coated sphere and an indium tin oxide (ITO) plate. It is shown that the illumination of an ITO plate by the laser pulses in ultra-violet (UV) leads to a dramatic decrease in the force magnitude. Possible explanation of this phenomenon is proposed. Measurements of the gradient of the Casimir force between a gold-coated sphere and a gold-coated plate and their comparison with different theoretical approaches are considered in Sec.4. In Sec.5 the Casimir effect between magnetic surfaces is considered. It is shown that the case of magnetic metals suggest new interesting opportunities for a selection between different theoretical approaches to the Casimir force. Special attention is paid to the Casimir effect for graphene in Sec.6. Unique properties of graphene lead to a unusual behavior of the Casimir force between two graphene sheets and between a graphene sheet and a plate made of some ordinary material. This allows modulation of the Casimir force by depositing graphene on material substrates. The results of first experiment on measuring the Casimir interaction between a gold sphere and a graphene-coated substrate are compared with the recently developed theory. In Sec.7 the reader will find our conclusions and discussion including the prospects for use of the Casimir effect in nanotechnology and microelectronics.

## 2. Modern achievements and challenges in the Casimir physics

To speak about modern achievements and problems, it is necessary to start with a brief formulation of the Lifshitz theory of dispersion forces [1-3].We consider two thick parallel plates (semispaces) at a separation $a$ at temperature T in thermal equilibrium with an environment. Let the materials of the first and second plates are characterized by the frequency-dependent dielectric permittivities $\varepsilon^{(1)}(\omega)$, $\varepsilon^{(2)}(\omega)$ and magnetic permiabilities $\mu^{(1)}(\omega)$, $\mu^{(2)}(\omega)$, respectively. Then the free energy of the fluctuating field per unit area of the plates (or the Casimir free energy) is given by [2, 3]

$$\mathcal{F}(a,T) = \frac{k_B T}{2\pi} {\sum_{l=0}^{\infty}}' \int_0^{\infty} k_\perp dk_\perp \sum_\alpha ln[1 - r_\alpha^{(1)}(i\xi_l, k_\perp) r_\alpha^{(2)}(i\xi_l, k_\perp) e^{-2aq_l}]. \quad (1)$$

Here, $k_B$ is the Boltzmann constant, $k\perp$ is the magnitude of the projection of the wave vector on the plane of plates, $\xi_l = 2\pi k_B T l/\hbar$ with $l = 0, 1, 2, \ldots$ are the Matsubara frequencies, the prime near the summation sign in $l$ adds the multiple

½ to the term with $l = 0$, $q_l = \sqrt{k_\perp^2 + \xi_l^2/c^2}$, and the sum in α means a summation over two independent polarizations of the electromagnetic field, transverse magnetic (α = TM) and transverse electric (α = TE). The reflection coefficients of the first and second plates (n = 1, 2, respectively) are the standard Fresnel coefficients of classical electrodynamics, but calculated at the imaginary Matsubara frequencies

$$r_{TM}^{(n)}(i\xi_l, k_\perp) = \frac{\varepsilon_l^{(n)} q_l - k_l^{(n)}}{\varepsilon_l^{(n)} q_l + k_l^{(n)}},$$

$$r_{TE}^{(n)}(i\xi_l, k_\perp) = \frac{\mu_l^{(n)} q_l - k_l^{(n)}}{\mu_l^{(n)} q_l + k_l^{(n)}},$$

(2)

where $\varepsilon_l^{(n)} = \varepsilon^{(n)}(i\xi_l)$, $\mu_l^{(n)} = \mu^{(n)}(i\xi_l)$ and

$$k_l^{(n)} = \sqrt{k_\perp^2 + \varepsilon_l^{(n)} \mu_l^{(n)} \frac{\xi_l^2}{c^2}}.$$

(3)

Thus, if we have the values of dielectric permittivities and magnetic permeabilities at the frequencies i$\xi_l$, we can also calculate the Casimir free energy by the Lifshitz formula (1). Then one can calculate the Casimir force per unit area of the plates (or the Casimir pressure) and the Casimir entropy per unit area

$$P(a,T) = -\frac{\partial \mathcal{F}(a,T)}{\partial a}, \quad S(a,T) = -\frac{\partial \mathcal{F}(a,T)}{\partial T}.$$

(4)

Using the proximity force approximation (PFA) valid for a large sphere of radius R in close proximity to a plate [2,10], it is possible also express the Casimir force acting in sphere-plate geometry

$$F(a,T) = 2\pi R \mathcal{F}(a,T).$$

(5)

This is an approximate equation. Recently it was shown [11,12] that its error is less than $a$/R (i.e., less than a fraction of a percent for typical experiments on measuring the Casimir force).

The major theoretical achievement of the last decade is the generalization of the Lifshitz formula (1) for the case of arbitrarily shaped test bodies [13-15].

It was achieved by using the scattering approach of quantum field theory and the formalism of functional determinants.

On the experimental side, a lot of measurements of the Casimir interaction have been performed in the configurations with metallic [16-19], semiconductor [20-27] and dielectric [28] surfaces. The measurement data of these experiments were compared with theoretical predictions of the Lifshitz theory and the results of this comparison turned out to be puzzling. Before we discuss this puzzle, it is necessary to consider the dielectric permittivities which should be substituted to the Lifshitz formula for different test bodies (here we deal with nonmagnetic test bodies, so that μ = 1; the magnetic properties are considered in Sec.5).

For the case of true dielectrics (insulators) with no free change carriers the dielectric permittivity at the imaginary Matsubara frequencies can be presented as a sum of contributions from the bound (core) electrons [1]

$$\varepsilon_c^{(n)}(i\xi_l) = 1 + \sum_{j=1}^{K} \frac{g_j^{(n)}}{\omega_j^{(n)2} + \xi_l^2 + \gamma_j^{(n)}\xi_l}, \qquad (6)$$

where $g_j^{(n)}$ are the oscillator strengths, $\omega_j^{(n)} \neq 0$ are the oscillator frequencies, $\gamma_j^{(n)}$ are the relaxation parameters, and K is the number of oscillators.

Each real dielectric at any nonzero temperature contains, however, some small fraction of free charge carriers and possesses the so-called dc conductivity $\sigma_0(T)$, i.e., the conductivity at a constant current. With account of dc conductivity, the dielectric permittivity of real dielectrics is given by

$$\varepsilon_d^{(n)}(i\xi_l) = \varepsilon_c^{(n)}(i\xi_l) + \frac{4\pi\sigma_0^{(n)}(T)}{\xi_l}, \qquad (7)$$

where the dielectric permittivity $\varepsilon_c^{(n)}$ determined by the core electrons is defined in (6).

Now we consider metals which contain high concentrations of free charge carriers (electrons). The dielectric permittivity of metals in the framework of the Drude model is given by

$$\varepsilon_D^{(n)}(i\xi_l) = \varepsilon_c^{(n)}(i\xi_l) + \frac{\omega_p^{(n)2}}{\xi_l[\xi_l + \gamma^{(n)}(T)]}, \qquad (8)$$

where $\omega_p^{(n)}$ and $\gamma^{(n)}(T)$ are the plasma frequencies and relaxation parameters of metals of the plates and $\varepsilon_c^{(n)}$ is again the contribution (6) due to core electrons. It is necessary to stress that the dielectric permittivities (7) and (8) along the real frequency axis $\omega = i\xi$ go to infinity as $\omega^{-1}$ when the frequency vanishes.

At sufficiently high frequencies (in the region of infrared optics) it holds $\xi_l \gg \gamma^{(n)}(T)$ and equation (8) turns into

$$\varepsilon_p^{(n)}(i\xi_l) = \varepsilon_c^{(n)}(i\xi_l) + \frac{\omega_p^{(n)2}}{\xi_l^2}, \qquad (9)$$

which is called the dielectric permittivity of the plasma model. In fact, the plasma model describes the free electron gas with no relaxation properties.

The first puzzling fact was revealed theoretically more than 10 years ago. It was proven analytically [29-31] that the Casimir entropy (4) calculated using the Lifshitz formula (1) and the dielectric permittivity of the Drude model (8) for metals with perfect crystal lattices goes to a nonzero limit

$$\lim_{T \to 0} S(a, T) = S_0\left(a, \omega_p^{(n)}\right) < 0, \qquad (10)$$

when the temperature vanishes. Taking into account that $S_0$ depends on the parameters of a system under consideration (such as the volume and the plasma frequencies), one arrives to the conclusion that the Casimir entropy violates the third law of the thermodynamics, the Nernst heat theorem [32]. Quite to the contrary, if one repeats the same calculation, but uses the dielectric permittivity of the plasma model (9), one arrives [29-31] to the zero entropy

$$\lim_{T \to 0} S(a, T) = 0 \qquad (11)$$

in accordance with the Nernst heat theorem.

It should be especially emphasized that the results (10) and (11) are entirely determined by the zero frequency ($l = 0$) contribution to the Lifshitz formula (1), i.e., by the low-frequency behavior of the dielectric permittivities. Thus, the violation of the Nernst theorem occurs when we use the Drude dielectric permittivity (8) applicable at low frequencies. At the same time, the Nernst theorem is satisfied when we use at low frequencies the plasma dielectric permittivity (9) which is applicable only at high frequencies, and this is puzzling.

A bit later, similar in spirit results were obtained for dielectric materials. It was shown [33-35] that the Casimir entropy, calculated using the Lifshitz formula (1) and the dielectric permittivity (7) with account the dc conductivity, goes to a nonzero limit

$$\lim_{T \to 0} S(a, T) = S_0 \left( a, \varepsilon_c^{(n)}(0) \right) > 0 \qquad (12)$$

when temperature goes to zero. If one calculates the Casimir entropy for two dielectric plates described by the dielectric permittivity (6) disregarding the dc conductivity, one arrives to the zero limit (11). Again, when the dielectric permittivity (7) of real dielectric plates is used, the Nernst heat theorem is violated. However, if one omits the dc conductivity in calculations of the Casimir entropy, the Nernst heat theorem is satisfied.

Quite unexpectedly for many experts in the field, these theoretical results were supported by several experiments made independently by three different experimental groups. In a series of experiments performed by R. S. Decca using a micromachined oscillator the effective Casimir pressure between two parallel gold plates has been measured [16-19]. In fact in these experiments the immediately measured quantity was not the Casimir pressure, but the gradient of the Casimir force acting between a gold-coated sphere and gold-coated plate, i.e., $\partial F(a,\text{T})/\partial a$. The latter quantity, however, can be easily recalculated into the Casimir pressure between two parallel plates, $P(a,\text{T})$, using the PFA (5)

$$P(a, T) = -\frac{1}{2\pi R} \frac{\partial F(a,T)}{\partial a}. \qquad (13)$$

The experimental data of experiments [16-19] have been compared with two theoretical approaches using the dielectric permittivity (8) (the Drude model approach) and using the dielectric permittivity (9) (the plasma model approach) at low frequencies. This was done in a conservative way with careful account of all experimental errors and uncertainties in theoretical parameters. As a result, it was found that the Drude model approach is excluded by the measurement data. The confidence level of this exclusion achieves 99% over some separation regions. The plasma model approach was found in agreement with the data at a 90% confidence level. Thus, the thermodynamically consistent approach was confirmed experimentally. The puzzle, however, has become even deeper because the experimental data excluded the dielectric permittivity (8), which is valid at low frequencies, and were found in agreement with the dielectric permittivity (9), which is not applicable at low frequencies. We remind that the relationship $\varepsilon(\omega) \sim \omega^{-1}$ at low, quasistatic, frequencies is a direct consequence of the Maxwell equations [36]. It was confirmed by thousands experiments in physics and electrical engineering during 150 years. We also stress that in the configuration of two gold test bodies it holds

$$\frac{\partial F_D(a,T)}{\partial a} < \frac{\partial F_p(a,T)}{\partial a}, \tag{14}$$

where $F_D$ and $F_p$ are the Casimir forces between a sphere and a plate calculated using the Drude and the plasma model approaches, respectively. We return to this inequality when discussing the case of magnetic metals in Sec.5.

Similar results were obtained in experiments with semiconductor and dielectric test bodies. U. Mohideen [26,27] measured the difference in the Casimir force acting between a Si membrane and a gold-coated sphere in the presence and in the absence of laser pulses on a membrane. Measurements were performed by means of the static AFM, so that the difference in the force was an immediately measured quantity. It is necessary to stress that in the absence of laser light the Si membrane was in a dielectric state. The Si was p-type doped with a relatively high nominal resistivity and low density of charge carriers $n \approx 5 \times 10^{14}$ cm$^{-3}$. In the presence of laser pulses on a membrane n was up to five orders of magnitudes higher, i.e., Si was in a metallic state. Thus, the measured quantity was

$$F_{diff}(a,T) = F_{mm}(a,T) - F_{dm}(a,T), \qquad (15)$$

where $F_{mm}$ is the Casimir force between two metallic test bodies (metallic Si plate and gold sphere) and $F_{dm}$ is the Casimir force between a dielectric Si plate and a gold sphere.

The difference force $F_{diff}$ was calculated using the Lifshitz theory and different dielectric permittivities (6) – (9). The calculation results were compared with the measurement data. It turned out that the data are not of sufficient precision to discriminate between the cases when metals are described by the Drude (8) or the plasma (9) models (this was already done in the experiments [16-19]). However, it was shown that the data exclude at the 95% confidence level the theoretical approach where the dielectric plate is described by the permittivity (7) taking into account the dc conductivity of a dielectric Si. The same data were found in agreement with theory disregarding the dc conductivity of a dielectric Si, i.e., using the permittivity (6). Thus, again, the measurement data were in agreement with the thermodynamically consistent approach. This agreement was, however, achieved at the expense of disregard by the free charge carriers which are present in any dielectric at any nonzero temperature.

One more experiment with dielectric test body is a measurement of the Casimir-Polder force between $^{87}$Rb atoms belonging to the Bose-Einstein condensate and a $SiO_2$ plate [28]. It was shown that the measurement data are in good agreement with theory when the dc conductivity of a $SiO_2$ plate is omitted in calculations [28]. The results of alternative calculation made with taken into account dc conductivity of $SiO_2$ [37] are excluded by the data. Again, the experiment supports thermodynamically consistent theory, but the free charge carriers in $SiO_2$ should be discarded for some unclear reason.

In the end of this section we briefly mention several other important results obtained during this period. Specifically, the lateral Casimir force between sinusoidally corrugated surfaces of a sphere and a plate was measured and found to be in a very good agreement with the generalization of the Lifshitz theory taking corrugations into account [38, 39]. The normal Casimir force between sinusoidally [40] and rectangular [41-44] corrugated surfaces was also measured and compared with theory. Several measurements and calculations of the Casimir interaction in fluids [45, 46] were performed. Much attention was paid to the diverse applications of the Casimir force in nanoscience and nanotechnology. Thus, the problem of stability in torsional nano-actuators was investigated [47, 48]. The Casimir force between two components of a silicon chip was demonstrated [49]. Furthermore, it was shown that the Casimir force in microdevices can be modulated and even canceled by using radiation pressure [50]. It was also proposed to use the lateral Casimir force for the frictionless transduction of motion in micromechanical systems [51, 52].

A more detailed information about these and other experiments can be found in the reviews [53-55].

### 3. New features of the Casimir force between a gold sphere and an indium tin oxide plate

The puzzle discussed in Sec.2 demonstrates that knowledge of the dielectric permittivities may be not sufficient to calculate the Casimir force in the framework of the Lifshitz theory. This statement is further illustrated by an experiment [56, 57] on measuring the Casimir force between a gold-coated sphere and an ITO plate before and after this plate undergoes the UV treatment.

Measurements were performed by means of an AFM schematically shown in Fig.1. Here, an Au-coated sphere of R = (101.2 ± 0.5) μm radius is attached to a cantilever which bends in response to the Casimir force acting between a sphere and a plate. This bending is detected by the deflection of the laser beam, leading to a difference signal between the photodiodes A and B. The plate is mounted on the top of the piezoelectric actuator which allows movement of the plate towards the sphere.

As was already shown previously, for metallic-type semiconductors like ITO, the Casimir force can be up to a factor of two smaller in magnitude than for good metals like gold [20-24]. A specific feature of the experiment [56,57] is that dielectric properties of ITO plates were measured by means of ellipsometry over a wide frequency range. What is more, measurements of the Casimir force between a sphere and a plate were performed for two times, i.e., for an untreated and UV-treated ITO plate. For this purpose, after the force measurements with an untreated plate were completed, it was placed in a special air chamber containing a UV lamp. This lamp emits a spectrum with the primary peak at the wavelength 254 nm and a secondary peak at 365 nm. During the UV treatment the sample was placed at 1 cm from the lamp for 12 h. After finish of the UV treatment, the force measurements were performed again. The ellipsometry measurements of the imaginary part of $\varepsilon(\omega)$ were performed for both untreated and UV-treated samples. It was shown that there are only minor differences for both samples. The obtained results for Im $\varepsilon(\omega)$ were recalculated into $\varepsilon(i\xi)$ using the Kramers-Kronig relation. It was demonstrated that the UV treatment does not lead to any significant changes in the dielectric permittivity of an ITO plate as a function of imaginary frequency. On this basis, following the standard Lifshitz theory, one could conclude that the measured Casimir forces between a gold sphere and either an untreated or UV-treated plate should be almost coinciding.

The measurement data, however, are against these expectations. In Fig.2 we show the measurement results for the mean Casimir forces as crosses, where the error bars are calculated at a 95% confidence level, for an untreated and UV-treated plate (lower and upper sets of crosses, respectively) at different plate-sphere separations. As can be concluded from the lower set of crosses (an untreated ITO plate) there is a 40% - 50% decrease in the force magnitude in comparison with the case of two gold bodies in agreement with previous work [23, 24]. However, if to compare the lower set of crosses with the upper set of crosses, it becomes clear that the magnitudes of the Casimir force from a UV-treated plate are 21% - 35% smaller than from an untreated plate. This is strange if to take into account that the dielectric permittivities in both cases are almost coinciding.

To get a better understanding of this situation, we compare the measurement data of Fig.2 with theory. The theoretical Casimir forces between a gold sphere and an untreated ITO plate are shown by the two lower solid lines enclosing the lower set of crosses. These lines are calculated by the Lifshitz theory using the dielectric permittivity of gold and of metallic ITO. Here, in the limits of experimental errors, the use of different extrapolations to low frequencies (by means of the Drude or plasma models) does not lead to large differences in the obtained results, but to only minor theoretical error. This error, and also a larger error due to uncertainty in the extrapolation of ellipsometry data to higher frequencies, lead to a theoretical band between the two solid lines. It is seen that the lower theoretical band is in good agreement with the measurement data for an untreated plate. It is also seen that the lower theoretical band is in complete disagreement with the upper set of crosses obtained for a UV-treated sample. These conclusions are in fact valid up to a separation of 200 nm [57].

To find possible explanation of this disagreement, we omit the contribution of free charge carriers in the case of UV-treated sample and extrapolate its dielectric permittivity along the imaginary frequency axis obtained from the ellipsometry measurements to low frequencies by a constant. The same procedure was employed in Sec.2 when we discussed the case of dielectric plates. Using this approach, we obtain the upper pair of the solid lines which enclose the set of crosses measured for a UV-treated sample. Thus, the experimental data for a UV-treated sample are in a good agreement with theory using the dielectric permittivity (6) which does not take free charge carriers into account.

At this point one can hypothesize [57] that the UV treatment of an ITO plate results in the Mott-Anderson phase transition [58] of metallic ITO to an insulator state without any noticeable change of its optical properties at room temperature. This hypothesis is supported by the observation that the UV treatment of ITO leads to a lower mobility of charge carriers [59]. The proposed hypothesis can be verified by the investigation of electrical properties of the UV-treated ITO at low temperatures. If the UV treatment really transforms the ITO film from metallic to dielectric state, the electric conductivity (which is similar for an untreated and UV-treated plates at room temperature) should vanish when the temperature vanishes. If, alternatively, the ITO plate remains in a metallic state after the UV treatment, the conductivity will go to a nonzero limiting value with decreasing temperature.

Basing on these ideas, a new experiment on measurement of the Casimir force or its gradient between a gold-coated sphere and two different plates made of doped semiconductors was proposed [60]. It was suggested that the concentration of charge carriers in one plate should be slightly below and in another plate slightly above the critical density at which the Mott-Anderson insulator-metal transition occurs [58]. This ensures that the dielectric permittivities of both plates at room temperature are almost identical so that minor differences between them cannot lead to large change in the magnitude of the Casimir force calculated using the standard Lifshitz theory. Taking into account already performed experiments [26-28, 56, 57], one may expect, however, that such a large change will be observed because in the dielectric state, against a literal application of the Lifshitz theory, one should discard the contribution of free charge carriers in order to get an agreement with the measurement data. If these ideas were confirmed, this would open opportunities for modifying the van der Waals and Casimir forces without change of the dielectric permittivity.

### 4. Measurement of the Casimir interaction between gold test bodies by means of dynamic atomic force microscope

In Sec.2 we have already discussed the experiments by R. S. Decca [16-19], where the gradient of the Casimir force between a gold sphere and a gold plate (and the effective Casimir pressure between two parallel gold plates) were measured by means of micromachined oscillator. Taking into account that the measurement data were challenging (the Drude model theoretical approach was excluded and the plasma model approach was confirmed), it is highly desirable to have an independent confirmation of these results obtained in another laboratory and using quite different experimental setup. Such an additional experiment was performed by U. Mohideen [61] using a dynamic AFM.

The general scheme of an AFM is already presented in Fig.1. The main difference of the dynamic regime used in [61] from the static regime used in [56, 57] is the following. The gold-coated sphere of radius $R = (41.3 \pm 0.2)$ μm was not static, but oscillating with a natural frequency of the oscillator $\omega_0$ in the vertical direction. The amplitude of oscillations was equal to only a few nanometers in order the oscillation system would stay in the linear regime. The Casimir force acting between a sphere and a plate causes a change in the resonance frequency $\Delta\omega(a) = \omega_r(a) - \omega_0$. In the linear regime this change is connected with the gradient of the Casimir force.

$$\Delta\omega(a) = -\frac{\omega_0}{2k}\frac{\partial F(a,T)}{\partial a}, \tag{16}$$

where k is the spring constant of the cantilever. Thus, by measuring the frequency shift of the oscillator, one obtains the gradient of the Casimir force as a function of separation between the sphere and the plate. Using the PFA (13) and the first equality in (4), the gradient of the Casimir force can be obtained theoretically from (1) in the framework of the Lifshitz theory

$$\frac{\partial F(a,T)}{\partial a} = 2k_B T R \sum_{l=0}^{\infty}{}' \int_0^{\infty} q_l k_\perp dk_\perp \sum_\alpha \frac{r_\alpha^{(1)}(i\xi_l, k_\perp) r_\alpha^{(2)}(i\xi_l, k_\perp)}{e^{2q_l a} - r_\alpha^{(1)}(i\xi_l, k_\perp) r_\alpha^{(2)}(i\xi_l, k_\perp)}. \qquad (17)$$

In this case of two gold test bodies $r_\alpha^{(1)} = r_\alpha^{(2)}$, but we preserve different reflection coefficients in (17) for application of this equation in the next sections.

Now we consider the measurement results for the mean gradients of the Casimir force and their comparison with theory. First of all, with the help of (13) it is possible to compare the mean gradients measured using the dynamic AFM in [61] with the mean Casimir pressures measured using the micromachined oscillator in [18,19]. This comparison demonstrates a perfect agreement of both sets of data in the limits of the experimental errors [61]. We further plot the mean measured gradients of the Casimir forces divided by the sphere radius as the upper sets of crosses in Figs. 3 and 4. The arms of the crosses represent the total experimental errors (i.e., the random and systematic combined) determined at a 67% confidence level. In fact, measurements were performed over a wider interval from 235 to 746 nm, but we consider the interval from 320 to 400 nm in order to make the data more visual. The lower sets of crosses in Figs. 3 and 4 are discussed in Sec.5. The upper solid line in Fig.3 presents the computational results for the gradient of the Casimir force obtained using (17) and the plasma model approach, i.e., with the help of dielectric permittivity (9) at low frequencies disregarding the relaxation properties of electrons. As discussed in Sec.2, this approach is consistent with thermodynamics, but uses the plasma model at low frequencies, where it is not applicable. In Fig.3 it is seen that the plasma model approach is in agreement with the data (in fact it is in agreement with the data over the entire measurement range from 235 to 746 nm [61]).

In Fig.4 the upper solid line presents the computational results obtained using (17) and the Drude model approach, i.e., by applying the dielectric permittivity (8) at low frequencies taking into account the relaxation properties of electrons. This approach violates the Nernst heat theorem, but respects the behavior of the dielectric permittivity at low frequencies as $\omega^{-1}$ in accordance to the Maxwell equations. As is seen in Fig.4, the Drude model approach is excluded by the measurement data. In fact, at the 67% confidence level the Drude model approach is excluded over the separation region from 235 to 420 nm. At the 95% confidence level it is excluded by the data over a more narrow separation region from 235 to 330 nm [62]. Thus, the results of earlier experiments [16-19] are independently confirmed.

Taking into account, however, that at separations of a few hundred nanometers the theoretical predictions of the Drude and plasma model approaches differ for only several percent, some doubts in the obtained results still remained. It is well known that there are the so-called patch potentials arising due to the polycrystal structure of metallic coatings or dust and contaminants on the surfaces. Such potentials may influence the measurement results of the Casimir force and their comparison with theory. This question was investigated in [17] using the conventional model of patches [63] and their role was shown to be negligibly small, but some authors developed other models of patches where they could contribute to the force significantly. It was even speculated [64] that surface patches may render the experimental data [18, 19] compatible with theoretical predictions of the Drude model approach. Thus, some decisive results are needed which would make the experimental situation fully transparent by excluding the role of any possible background effect. Such results were obtained very recently by measuring the Casimir interaction between two ferromagnetic surfaces [62, 65] and between a nonmagnetic and magnetic test bodies [66]. As is shown in the next section, the use of magnetic surfaces provides a possibility to tremendously increase the difference in theoretical predictions of the Drude and plasma model approaches to calculation of the Casimir force.

5. Investigation of the Casimir effect for magnetic test bodies

In two successive experiments performed by means of dynamic AFM, the gradient of the Casimir force was measured between a gold sphere and a Ni plate [66] and between two Ni surfaces of a sphere and a plate [62, 65]. The ferromagnetic metal Ni is characterized by the static magnetic permeability $\mu_0 = \mu(0) = 110$ and had no spontaneous magnetization. It is well known [67] that for ferromagnetic metals $\mu(i\xi)$ becomes equal to unity at $\xi > 10^5$ Hz. Taking into account that at room temperature the first Matsubara frequency $\xi_1 \sim 10^{14}$ Hz, one can conclude [68] that the magnetic Casimir interaction is determined by only the zero-frequency term, $l = 0$, in the Lifshitz formulas (1) and (17). Thus, in all computations of the Casimir interaction for magnetic surfaces, one can put $\mu = \mu_0$ at $l = 0$ and $\mu(i\xi_l) = 1$ at all $l \geq 1$.

Measurements of the gradient of the Casimir force between a gold sphere and a Ni plate [66] and between a Ni sphere and a Ni plate [62, 65] were made in the same way and using the same setup, as described in previous section for the case of two gold test bodies. As to theoretical calculations, for magnetic test bodies the most important zero-frequency contributions to the Lifshitz formula are different from the case of nonmagnetic metals. This difference is absent for the TM polarization of the electromagnetic field. Really, from (2) it is seen that

$$r_{TM}^{(n)}(0, k_\perp) = 1 \qquad (18)$$

for both magnetic and nonmagnetic metals and for both models of the dielectric permittivity (8) and (9). For the TE reflection coefficient in (2) the situation is more complicated. If the Drude model dielectric permittivity (8) is used one obtains

$$r_{TE}^{(n)}(0, k_\perp) = 0 \quad \text{or} \quad r_{TE}^{(n)}(0, k_\perp) = \frac{\mu_0^{(n)} - 1}{\mu_0^{(n)} + 1} \qquad (19)$$

for nonmagnetic and magnetic metals, respectively. Thus, in this case the results are qualitatively different. If the plasma model (9) is used we have instead

$$r_{TE}^{(n)}(0, k_\perp) = \frac{\mu_0^{(n)} k_\perp - \sqrt{k_\perp^2 + \mu_0^{(n)} \omega_p^{(n)2}/c^2}}{\mu_0^{(n)} k_\perp + \sqrt{k_\perp^2 + \mu_0^{(n)} \omega_p^{(n)2}/c^2}}. \tag{20}$$

This expression, as different from (19), provides a nonzero result for both nonmagnetic and magnetic metals. Recently it was proven that for magnetic metals described by the plasma model the Nernst heat theorem is satisfied. If the Drude model is used to describe the dielectric permittivity of magnetic metals, the Nernst heat theorem is violated [69]. Similar statements were proven also for magnetic dielectric materials. Specifically, the Nernst heat theorem for these materials is satisfied or violated if one uses the dielectric permittivities (6) or (7), respectively [70].

Now we consider the measurement results and their comparison with theory. The measured gradients of the Casimir force between two Ni surfaces of a sphere and a plate divided by the sphere radius are shown by the lower sets of crosses in Figs.3 and 4, where the error bars are determined at the 67% confidence level. For this experiment [62, 65] the sphere radius $R = 61.71 \pm 0.09$ μm. By comparing the upper and lower sets of crosses in Fig. 3 and 4, one can conclude that magnetic properties influence the value of the Casimir force. The calculation results using the plasma model approach are shown by the lower solid line in Fig. 3. It is seen that the measurement data are in a very good agreement with theory disregarding the relaxation properties of free electrons. The same statement is valid over the entire measurement range in this experiment from 223 to 550 nm.

The lower solid line in Fig.4 demonstrates the calculation results using the dielectric permittivity of the Drude model (8). It is seen that the Drude model approach is excluded by the measurement data. The analysis of the data and the calculation results over the entire measurement range shows that at the 67 % confidence level the Drude model approach is excluded in the separation range from 223 to 420 nm, and at the 95 % confidence level in the separation range from 223 to 345 nm [62].

As is seen in Figs. 3 and 4, for magnetic metals the predictions of the Lifshitz theory using the plasma and Drude model approaches satisfy an inequality

$$\frac{\partial F_p(a,T)}{\partial a} < \frac{\partial F_D(a,T)}{\partial a} \ . \tag{21}$$

This is just the opposite to an inequality (14) found for the first time in [16-19] for the case of nonmagnetic metals (Figs. 3 and 4 illustrate the inequality (14) as well).

The inequalities (14) and (21) taken together lead to important conclusions concerning the possibility of some unaccounted systematic effect (for instance, due to surface patches discussed in the end of Sec. 4) which could influence the results of the Casimir force measurements. If to admit that an additional attractive force due to surface patches brings the experimental date for two gold test bodies in agreement with the predictions of the Drude model approach, it would only increase the disagreement of the Drude model approach with the data for two Ni test bodies (leading also to a disagreement of the plasma model approach with the same data). It is not logical, however, to assume that the patch effect plays the role for gold, but does not play the same role for Ni. Thus, it is confirmed that the effect of patches is sufficiently small and does not play any important role in the measurements of the Casimir force.

This conclusion finds one more confirmation in the experiment on measuring the gradient of the Casimir force between a gold sphere and a Ni plate [66]. In this configuration at separations up to a few hundred nanometers, the theoretical predictions of the Drude and plasma model approaches almost coincide. The common prediction of both approaches was found to be in a very good agreement with the measurement data over the entire range of separations. Because of this, any additional force, either attractive or repulsive, would bring the data in disagreement with both theoretical approaches. Note also that recently the effect of surface patches for the test bodies used in Casimir experiments was directly measured by means of Kelvin probe microscopy and found to be negligibly small [71].

In spite of the strong support, which is given by the experiments with magnetic test bodies [62, 63, 65] to the plasma model approach, the differences between the predictions of both competing approaches in these experiments at separations below one micrometer remain to be of about a few percent. The use of magnetic test bodies allows, however, to suggest the experimental scheme, where the theoretical predictions of the Drude and plasma model approaches for the measured quantity differ by a factor of 1000 [72]. Such a scheme consists of two neighboring strips, one of which is made of magnetic metal Ni and the other one of nonmagnetic metal gold, and both strips are covered by an overlayer of gold of a few tens nanometer thickness [73]. The Ni sphere of about 150 μm radius scans across the hidden border of gold and Ni strips back and forth at a constant height. As a result, the immediately measured quantity is the difference of the Casimir forces, one between a Ni sphere and a gold overlayer above a Ni strip, and another one between a Ni sphere and a gold plate [73]:

$$F_{diff}(a,T) = F_{NiAuNi}(a,T) - F_{NiAu}(a,T). \qquad (22)$$

Simple calculation shows that in the framework of the Drude model approach $F^D_{diff} \sim 10^{-12}$ N, whereas in the framework of the plasma model approach $F^p_{diff} \sim 10^{-15}$ N. So large difference in theoretical predictions of the two approaches is achieved due to equations (19) and (20). When the Drude model approach is used, the TE mode passes through the gold overlayer and is reflected on the surface of a Ni strip, but not on a gold strip. In the case of the plasma model, there is no so big difference in reflections of the TE mode on both strips. As a result, the quantity (22) is much smaller. The first runs of this experiment have already been performed by R. S. Decca [73]. The data are found to be consistent with the plasma model approach and exclude the Drude model approach [73].

6. The Casimir effect for graphene

Graphene is a two-dimensional sheet of carbon atoms which is a prospective material for nanotechnology due to its unusual electrical, mechanical and optical properties [74]. As an element of microdevices, graphene can be situated at distances of the order of hundreds and even tens nanometers from the other elements. At these distances the van der Waals and Casimir forces become dominant. This is the reason why the Casimir effect for graphene attracts much attention in the literature.

At low energies, graphene is described by means of the Dirac model, which assumes the linear dispersion relation for massless quasiparticles moving with the Fermi velocity rather than with the velocity of light [75]. Using this model, a number of studies of graphene-graphene and graphene-material plate Casimir interactions were undertaken. Thus, the van der Waals coefficient for two graphene sheets at zero temperature was calculated using the correlation energy from the random-phase approximation [76, 77]. The Casimir force between two graphene sheets and between a graphene and a material plate was expressed via the Coulomb coupling of density fluctuations [78]. It was shown [78] that for graphene the thermal effects become crucial at much shorter separations than for all ordinary, three-dimensional, materials considered an previous sections. The graphene-graphene Casimir interaction was computed under an assumption that the conductivity of graphene can be described by the in-plane optical properties of graphite [79]. It was shown that for a sufficiently large mass-gap parameter of graphene the thermal Casimir force can vary significantly with varying temperature [80]. The reflection coefficients on graphene deposited on material substrates were expressed via the dielectric permittivity of a substrate and either a conductivity or a density – density correlation function of graphene [81-83]. It should be taken into account, however, that neither conductivity of graphene nor the density-density correlation function were known at nonzero temperature. Only some partial results were available (e.g., the longitudinal conductivity, respectively, the longitudinal density-density correlation function in the local approximation).

An important step towards the complete theory of Casimir forces in graphene systems was made in papers [84, 85], where the polarization tensor of graphene in (2+1) - dimensional space-time was found at zero and nonzero temperature. In terms of the polarization tensor $\Pi_{ik}(i\xi_l, k\perp)$, the exact reflection coefficients on graphene in the framework of the Dirac model are given by [85]

$$r_{TM}^{(g)}(i\xi_l, k_\perp) = \frac{q_l \Pi_{00}(i\xi_l, k_\perp)}{2\hbar k_\perp^2 + q_l \Pi_{00}(i\xi_l, k_\perp)},$$

$$(23)$$

$$r_{TE}^{(g)}(i\xi_l, k_\perp) = -\frac{k_\perp^2 \Pi_{tr}(i\xi_l, k_\perp) - q_l^2 \Pi_{00}(i\xi_l, k_\perp)}{2\hbar k_\perp^2 q_l + k_\perp^2 \Pi_{tr}(i\xi_l, k_\perp) - q_l^2 \Pi_{00}(i\xi_l, k_\perp)},$$

where tr denotes the sum of diagonal components $\Pi_k^{\ k}$.

Using the polarization tensor, the detailed computations of graphene-graphene [86] and graphene-real metal [87] Casimir interactions have been performed, and the Casimir-Polder interaction between different atoms and graphene sheet has been studied [88]. This allowed also calculation of the classical Casimir and Casimir-Polder interactions with graphene which hold when the separation distances are sufficiently large [89, 90].

For applications to nanotechnology, it is important to investigate the Casimir interaction with graphene deposited on different material substrates. To solve this problem, it is necessary to find the reflection coefficients on a two-layer system where one of the layers in made of some usual material described by the frequency-dependent dielectric permittivity and another one (graphene) is two-dimensional and is described by the polarization tensor. This problem was solved in 2014 using two different approaches. In the first approach [91], an exact equivalence between the polarization tensor and the density-density correlation functions of graphene was established. This allowed to express the previously known reflection coefficients [83] on the graphene-coated substrates in terms of the polarization tensor. In the second approach [92], the reflection coefficients on graphene-coated substrates were directly expressed via the polarization tensor using the formalism of multiple reflections. The results of both approaches are coinciding. They can be presented in the form [92]

$$r_{TM}^{(g,s)}(i\xi_l, k_\perp) = \frac{\varepsilon_l q_l + k_l(\frac{q_l}{\hbar k_\perp^2}\Pi_{00} - 1)}{\varepsilon_l q_l + k_l(\frac{q_l}{\hbar k_\perp^2}\Pi_{00} + 1)},$$

$$(24)$$

$$r_{TE}^{(g,s)}(i\xi_l, k_\perp) = \frac{q_l - k_l - \frac{1}{\hbar k_\perp^2}(k_\perp^2\Pi_{\mathrm{tr}} - q_l^2\Pi_{00})}{q_l + k_l + \frac{1}{\hbar k_\perp^2}(k_\perp^2\Pi_{\mathrm{tr}} - q_l^2\Pi_{00})},$$

where $\varepsilon_l = \varepsilon(i\xi_l)$ is the dielectric permittivity of a substrate.

This opened opportunities for a comparison between calculations and measurements of the Casimir interaction for graphene. The first experiment on measuring the Casimir interaction in graphene systems was performed for a gold-coated sphere of $R = 54.1$ μm radius and graphene deposited on a $SiO_2$ film covering a Si plate [93]. The gradient of the Casimir force in this system was measured using a dynamic AFM and the same experimental procedures, as described in Sec.4. In Fig.5 the mean measured gradients of the Casimir force are shown as crosses at different separations, where all errors are determined at the 67% confidence level [92]. Measurements were performed over the separation region from 224 to 500 nm. Taking into account the role of graphene for numerous applications, in Fig.5 we reproduce the measurement results over the entire measurement range.

The gradients of the Casimir force in the configuration of an experiment were computed by Eq. (17) using additionally modified reflection coefficients similar to (24), but with account of the fact that not one, but two material subtracts, $SiO_2$ and Si, have been used. The computed gradients of the Casimir force are shown as gray bands in Figs.5(a) − 5(d). The widths of the bands are determined by the uncertainty in the plasma frequency for Si, differences between the predictions of the Drude and plasma model extrapolations of the dielectric permittivities for gold and Si, and by the uncertainty of the mass gap parameter for graphene (only the pristine, perfect, graphene is gapless, but for real graphene deposited on a substrate, as in the experiment, quasiparticles may have a small but nonzero mass).

As is seen in Fig.5, the theoretical results computed using the Lifshitz theory and the reflection coefficients (24) are in a very good agreement with the measured gradients of the Casimir force. This opens opportunities for depositing graphene sheets on different substrates in order to modify the Casimir force in a predictable way.

Now we illustrate [94] the influence of graphene coating on the Casimir pressure between two parallel plates made of different materials, both metallic and dielectric, at $T = 300$ K. In Fig.6 we present the computational results for the ratios $P_{gg}/P$, where $P_{gg}$ is the pressure between two coated and $P$ between uncoated plates, as functions of separation. The solid lines from bottom to top are plotted for the plates made of gold, high-resistivity Si, sapphire, mica and fused silica, respectively. As seen in Fig.6, the Casimir pressure between gold plates is not affected by graphene coating over the range of separations from 100 nm to 6 μm. For dielectric plates graphene coatings result in a strong impact on the Casimir pressure. As can be seen in Fig.6, this impact increases with decreasing static dielectric permittivity of the plate material (we remind that for high-resistivity Si, sapphire, mica and fused silica the static dielectric permittivities are equal to 11.7, 10.1, 5.4, and 3.8, respectively). Thus, for two graphene-coated fused silica plates one obtains $P_{gg}/P = 1.47$, 1.72, 2.28, and 3.34 at separation distances 200 nm, 400 nm, 1 μm, and 6 μm, respectively.

The formalism of the polarization tensor is very effective not only in calculations of the Casimir force between graphene sheets, but enables one to solve the problem of the reflectivity properties of graphene. Previously the reflectivity of graphene was investigated using the local model for its conductivity and only some partial results for the reflectivity at low [81] and high [82] frequencies were obtained for the TM polarized light. The polarization tensor in the representation [85] was recently applied to calculate the reflectivity of graphene and graphene-coated substrates at high (optical) frequencies [95]. In so doing, the results of [82] for the TM polarized light were reproduced. It was shown, however, that with increasing angle of incidence the TE reflectivity of graphene monotonously increases. This is in disagreement with previous qualitative result that the TE reflectivity does not depend on the angle of incidence. In Fig.7 the TM and TE reflectivities of graphene multiplied by the factor $10^4$ are plotted as functions of the incidence angle $\theta_i$ in the case of high frequencies. As is seen in Fig.7, the TM and TE reflectivities are equal only at the normal incidence. With increasing angle of incidence, the TE reflectivity increases, whereas the TM reflectivity decreases monotonously.

The representation for the polarization tensor [85] does not admit analytic continuation to the entire real frequency axis and cannot be used to describe the reflectivities of graphene at all frequencies at nonzero temperature. This problem was solved very recently [96] by obtaining an alternative representation for the polarization tensor which coincides with that of [85] at all Matsubara frequencies, but, in contrast to it, satisfies all physical requirements along the real frequency axis. The use of new representation for the polarization tensor solves the problem of the reflectivity properties of graphene and opens further opportunities for investigation and application of this prospective material.

## 7. Conclusions and Discussion

In the foregoing, we have discussed recent advances and problems in physics of the van der Waals and Casimir forces. It was underlined that these forces are of common fluctuation origin and the use of different names is a matter of convention. We have considered the most precise experiments on measuring the Casimir interaction and comparison of the measurement data with theoretical predictions of the Lifshitz theory. It turns out that there is a fundamental unresolved problem in theory-experiment comparison. For metallic test bodies the Lifshitz theory is in agreement with the data only under a condition that the relaxation properties of condition electrons are omitted in computations. For dielectric test bodies the experimental data agree with theoretical predictions if the role of dc conductivity is disregarded. It is a remarkable fact that the experimentally consistent theoretical approaches are in agreement with thermodynamics. It is not a satisfactory situation, however, that an agreement with the data is achieved by disregarding well familiar relaxation properties at low frequencies and is replaced with a clear disagreement when these properties are taken into account. The problem calls for further investigation and promises breakthrough results in near future touching the foundations of quantum statistical physics.

As was mentioned in Sec.2, the van der Waals and Casimir forces find diverse applications in nanotechnology, including nano-actuators and various microdevices. This is because at short separations below a micrometer the Casimir force becomes larger in magnitude than the characteristic electric forces and determines the functionality of a microdevice. With further miniaturization, which is the main tendency of modern nanotechnology, the role of the Casimir force should inevitably increase. As mentioned in Sec.2, the Casimir force opens opportunities for the frictionless transduction of motion in micromechanical devices [51, 52]. The problems of tribology on microscales await for their resolution with the use of repulsive Casimir force predicted for different nanomaterials [97] and ferromagnetic dielectrics [68]. The Casimir force has already found applications in microswitches [98, 99] and is investigated in order to avoid pull-in and stiction in microdevices [100]. As was argued in Sec.6, further investigations and applications of graphene and other carbon nano-structures are also closely connected with the Casimir physics. If to take into account also that experiments on measuring the Casimir force are used to obtain stronger constrains on the corrections to Newton's law of gravitation [16-19, 101] and on the axion as a probable constituent of dark matter in astrophysics [102, 103], the prime importance of this subject for both fundamental science and its technological applications becomes quite evident.

Fig.1. Schematic diagram of the experimental setup for measuring the Casimir force by means of an atomic force microscope (see text for further discussion).

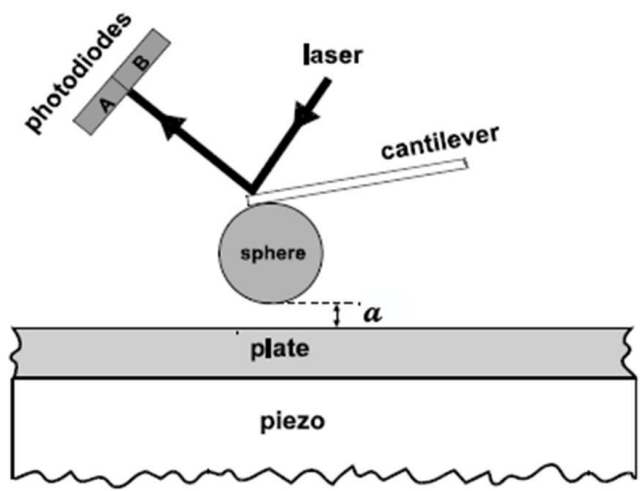

Fig.2. The mean measured Casimir forces between a gold sphere and ITO plate as functions of separation are indicated as lower and upper sets of crosses plotted at the 95% confidence level for the untreated and UV-treated samples, respectively. The respective lower and upper pairs of the solid lines show the theoretical results computed with included and omitted contributions of free charge carriers.

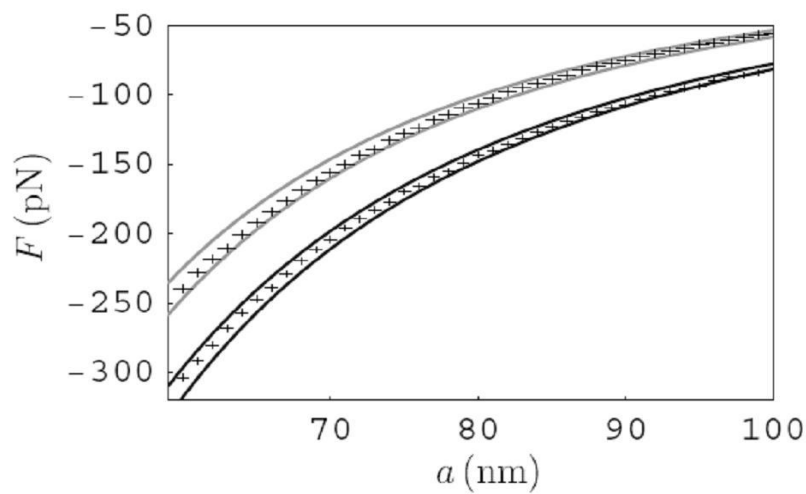

Fig.3. The mean measured gradients of the Casimir force divided by the sphere radii as functions of separation are indicated as upper and lower sets of crosses plotted at the 67% confidence level for two gold surfaces of a sphere and a plate, and for two Ni surfaces of a sphere and a plate, respectively. The respective upper and lower solid lines show the theoretical results computed using the plasma model approach.

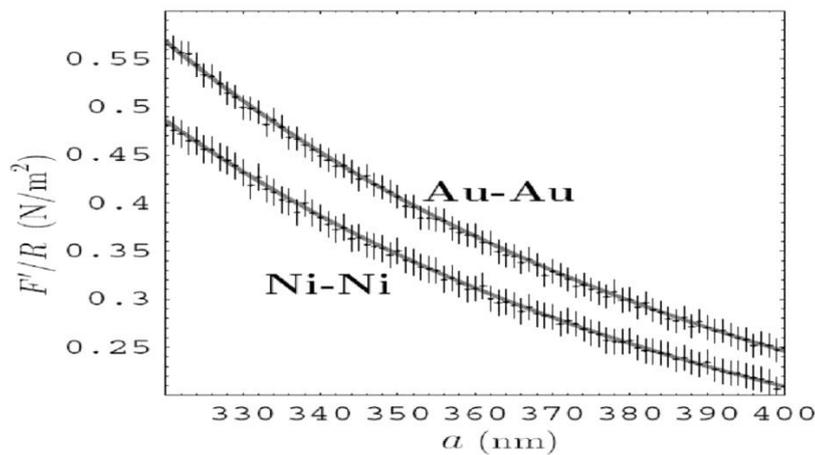

Fig.4. The mean measured gradients of the Casimir force divided by the sphere radii as functions of separation are indicated as upper and lower sets of crosses plotted at the 67% confidence level for two gold surfaces of a sphere and a plate, and for two Ni surfaces of a sphere and a plate, respectively. The respective upper and lower solid lines show the theoretical results computed using the Drude model approach.

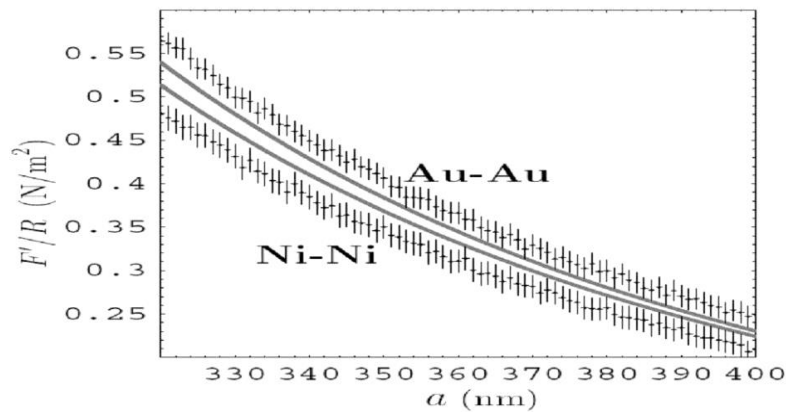

Fig.5. The mean measured gradients of the Casimir force between a gold sphere and graphene sheet deposited on a silica film covering a Si plate are indicated as crosses plotted at the 67% confidence level. The theoretical force gradients computed using the polarization tensor of graphene and dielectric permittivities of silica and Si are shown as the gray bands. The subfigures a-d demonstrate different ranges of separation.

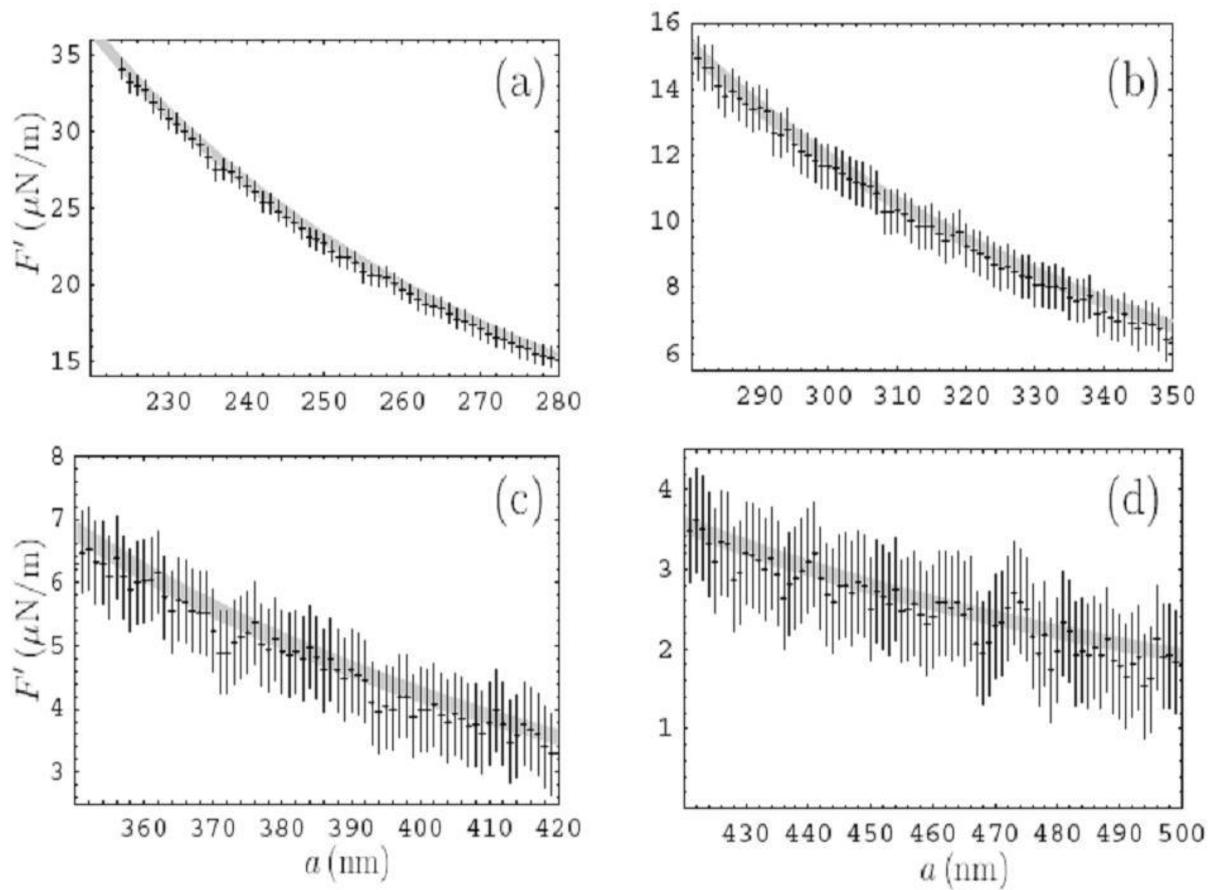

Fig.6. Ratios of the Casimir pressures between two plates coated with graphene to the Casimir pressures between uncoated plates made of the same material are shown as functions of separation.The lines from bottom to top are for the plates made of gold, Si, sapphire, mica, and fused silica, respectively.

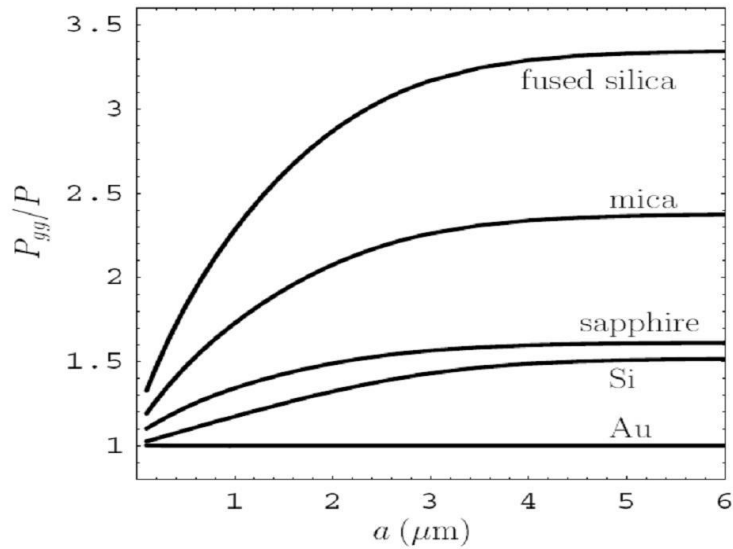

Fig. 7. Reflectivities of the transverse magnetic and transverse electric electromagnetic waves of optical frequencies on graphene are shown as functions of the angle of incidence by the lower and upper lines, respectively.

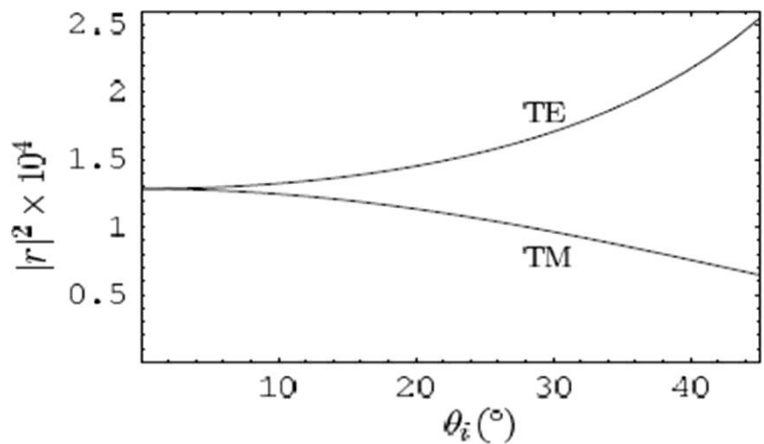